\documentclass[12pt]{article}
\usepackage{amsmath,amssymb,bm,epsf,epsfig,graphicx}
\input epsf.sty
\newcommand{\re}{\mathop{\rm Re}\nolimits}

\def\aver#1{\left\langle\, #1 \,\right\rangle}

\newcommand{\VEV}[1]{\left\langle #1\right\rangle}

\def \be {\begin{equation}}
\def \ee {\end{equation}}
\def \bea {\begin{eqnarray}}
\def \eea {\end{eqnarray}}
\def \bdm {\begin{displaymath}}
\def \edm {\end{displaymath}}

\begin{document}
{}~ \hfill\vbox{\hbox{YITP-11-29} }\break
\vskip 2.1cm

\centerline{\large \bf On Multibrane Solutions in Open String Field Theory}
\vspace*{8.0ex}

\centerline{\large \rm Masaki Murata$^{a,b,}$\footnote{Email: {\tt masaki at yukawa.kyoto-u.ac.jp}},
Martin Schnabl$^{a,}$\footnote{Email: {\tt schnabl.martin at gmail.com}}
}

\vspace*{8.0ex}

\centerline {\large \it $^a$ Institute of Physics AS CR, Na Slovance 2, Prague 8, Czech Republic}
\vspace*{2.0ex}

\centerline {\large \it $^b$ Yukawa Institute for Theoretical Physics, Kyoto University,}
\vspace*{2.0ex}
\centerline {\large \it Kyoto 606-8502, Japan}
\vspace*{2.0ex}

\vspace*{6.0ex}

\centerline{\bf Abstract}
\bigskip
Existence of open string field theory solutions describing configurations of multiple space-filling D-branes has been a subject of numerous speculations for quite some time. In this talk we present some new results
giving further support to these ideas.

\vfill \eject

\baselineskip=16pt


\section{Introduction}

Since the original insight by Sen \cite{SenCon} that the tachyon vacuum of open string field theory can be constructed in the so-called universal sector of the conformal field theory, and that it describes a state with no D-brane, it led naturally to the question whether there are other physically distinct solutions in this sector. In particular, whether there is a solution describing a state of multiple D-branes. The result of numerous numerical searches in level truncation was inconclusive if not negative, but there were good arguments based on vacuum string field theory which suggested otherwise \cite{RSZ, GT1}.

The starting point of the present work was a simple observation, that up to this date, there had been no computation of the energy for a class of analytic solutions of the open string field theory of the form\footnote{The notation used here is identical to that of many recent papers on the subject, see e.g. \cite{lightning} for a lightning review of the subject.}
\be\label{Fsol}
\Psi = F c \frac{K B}{1-F^2} c F,
\ee
where $F(K)$ is an arbitrary analytic function of $K$. In this form the solution was written down first by Okawa \cite{Okawa} generalizing second author's original solution \cite{Schnabl}  corresponding to the choice $F(K) = e^{-K/2}$.

The outcome of the energy computation should be a functional that given an analytic function $F$ produces a number. A very distinctive feature of this functional is that it has to be invariant under smooth deformations of $F$, such that $(1-F^2)^{-1}$ is also smooth, since arbitrary value $F'$ can be obtained from $F$ via a gauge transformation \cite{ErS}
$\Psi \to \Psi'= U^{-1} (Q+\Psi) U$, where
\bea
U &=& 1- F Bc F + \frac{1-F^2}{1-F'^2} F' B c F', \nonumber\\
U^{-1} &=& 1- F' Bc F' + \frac{1-F'^2}{1-F^2} F B c F.
\eea
Although this might be a sufficient information for making educated guesses, let us instead proceed with a careful computation.

For the string field to be geometric (i.e. a superposition of wedge states with insertions, see \cite{lightning} for further discussion) we demand that
\be
F(K) = \int_0^\infty d\alpha f(\alpha) e^{-\alpha K}.
\ee
This imposes the aforementioned condition of analyticity in $K$. Very reasonable additional condition is called $L_0$-safety which is equivalent to the statement that $F(z)$ be holomorphic for $\re z >0$ and bounded by a polynomial in $|z|$ in the whole domain $\re z \ge 0$.

\section{Energy computation}
The energy of a general solution of the form (\ref{Fsol}) can be most easily computed from the kinetic term. After few simple algebraic manipulations one finds
\bea\label{psiQpsi}
\aver{\Psi, Q\Psi} &=& \aver{\frac{K}{G}, (1-G), \frac{K}{G}, K G} - \aver{K, (1-G), \frac{K}{G}, K } \\\nonumber && - \aver{\frac{K}{G}, (1-G), K, K } + \aver{K, (1-G), K,\frac{K}{G} },
\eea
where
\be
G = 1- F^2
\ee
and to simplify our notation we have introduced
\be\label{gen_cor}
\aver{F_1, F_2, F_3, F_4} =
\aver{F_1(K) c F_2(K) c F_3(K) c F_4(K) c B}
\ee
for general $F_i(K)$. Assuming that all $F_i$ are $L_0$-safe and geometric, we can compute this basic correlator by writing
\be
F_i(K) = \int_0^\infty f_i(\alpha) e^{-\alpha K},
\ee
and using a formula from \cite{Schnabl,Okawa,Erler}
\bea\label{exp_cor}
\VEV{e^{-\alpha_1K}ce^{-\alpha_2K}ce^{-\alpha_3K}ce^{-\alpha_4K}cB}
&=& \VEV{c(\alpha_1)c(\alpha_1+\alpha_2)
c(\alpha_1+\alpha_2+\alpha_3)c(\alpha_1+\alpha_2+\alpha_3+\alpha_4)B}_{C_s}\nonumber\\
&=&  \frac{s^2}{4\pi^3}\Bigg[
\alpha_4\sin\frac{2\pi\alpha_2}s - (\alpha_3+\alpha_4)\sin\frac{2\pi(\alpha_2+\alpha_3)}s\nonumber\\
&&
+ \alpha_2\sin\frac{2\pi\alpha_4}s - (\alpha_2+\alpha_3)\sin\frac{2\pi(\alpha_3+\alpha_4)}s\nonumber\\
&&
+ \alpha_3\sin\frac{2\pi(\alpha_2+\alpha_3+\alpha_4)}s
+ (\alpha_2+\alpha_3+\alpha_4)\sin\frac{2\pi\alpha_3}s\Bigg],\nonumber
\label{eq:cF4}
\eea
where we have denoted $s= \sum_{i=1}^4 \alpha_i$. To find the energy one needs
the quadruple integral
\be
\int_0^\infty\!\!\int_0^\infty\!\!\int_0^\infty\!\!\int_0^\infty\! d\alpha_1 d\alpha_2 d\alpha_3 d\alpha_4 \, f_1(\alpha_1) f_2(\alpha_2) f_3(\alpha_3) f_4(\alpha_4) \VEV{e^{-\alpha_1K}ce^{-\alpha_2K}ce^{-\alpha_3K}ce^{-\alpha_4K}c B},
\ee
which in general might be quite hard to evaluate, especially since $f_i$ will typically be distributions and not ordinary functions. It is thus desirable to reexpress the correlator (\ref{gen_cor}) in terms of $F_i$ only. To do that, we use a little trick. Inside the multiple integral over $\alpha_i$ we insert an identity in the form
\be
1 = \int_0^\infty ds \, \delta \left(s-\sum_{i=1}^4 \alpha_i \right) = \int_0^\infty ds \int_{-i \infty}^{+i \infty} \frac{dz}{2\pi i}  \,\, e^{sz} \, e^{-z\sum_{i=1}^4 \alpha_i},
\ee
which allows us to treat $s$ as independent of the other integration variables $\alpha_i$. The integral over all $\alpha_i$ can be easily performed and we find
\bea\label{F1234}
\aver{F_1, F_2, F_3, F_4} &=&\int_0^\infty \! ds \int_{-i \infty}^{i \infty} \frac{dz}{2\pi i} \, \frac{s^2}{4\pi^3} e^{s z} \frac{1}{2i} \biggl[- F_1 \Delta F_2 F_3F_4' + F_1 \Delta(F_2 F_3') F_4 \nonumber\\
&& \qquad + F_1 \Delta(F_2 F_3) F_4' - F_1 F_2' F_3 \Delta F_4 +F_1 F_2 \Delta(F_3' F_4) +F_1 F_2' \Delta(F_3 F_4)
\nonumber\\
&& \qquad + \Delta F_1 F_2 F_3' F_4 + (s F_1+F_1')F_2 \Delta F_3 F_4 \biggr],
\eea
where for convenience we have omitted common argument $z$ and also introduced an operator $\Delta$ defined as
\be
(\Delta F)(z) = F\left(z-\frac{2\pi i}{s}\right) - F\left(z+\frac{2\pi i}{s}\right).
\ee
Using this general formula to evaluate (\ref{psiQpsi}) one finds long expression with 38 terms. To simplify this further we can close the integration contours by adding a half-circle at infinity in the $\re z<0$ half-plane. If the falloff of the integrand is sufficiently fast, these extra contributions are actually vanishing. Most of the terms depend on $z$ through the combination $z \pm \frac{2\pi i}{s}$. Having closed the contour, one can now split the integral into 38 individual integrals, apply change of variables and deform the contour back to its original position for each of the terms. With a judicious choice of these shifts which removes $s$ dependence of the denominators the whole expression simplifies to\footnote{We have performed the algebraic manipulations symbolically with Mathematica.}
\be
\aver{\Psi, Q \Psi} = \frac{3}{\pi^2} \int_0^\infty\!\! ds \oint\!\! \frac{dz}{2\pi i}\, e^{s z} \left[z^2 s \frac{G'(z)}{G(z)}+
(z\partial_z-s\partial_s)\left(\frac{i z s^2}{8\pi} \frac{\Delta(z G)(z)}{G(z)}\right) \right].
\ee
The second term is effectively a total derivative which under suitable conditions does not contribute, while the first term can be simplified by performing the $s$ integral.
The final result for the energy $E= \frac{1}{6} \aver{\Psi, Q \Psi}$ is thus
\be\label{resA}
\frac{E}{\rm Vol} = \frac{1}{2\pi^2} \oint_{C}\!\! \frac{dz}{2\pi i}\, \frac{G'(z)}{G(z)},
\ee
where $C$ is a contour running upward the imaginary axis, bypassing possible singularity at the origin on the left, and finally closing at infinity in the $\re z<0$ plane. The open string coupling constant has been set to unity and ${\rm Vol}$ denotes the D-brane volume.

\section{Energy from the closed string overlap}

Ellwood has recently observed \cite{Ellwood} that the value of Shapiro-Thorn-Hashimoto-Itzhaki-Gaiotto-Rastelli-Sen-Zwiebach invariant, or Ellwood invariant for short,
\be
\aver{I|\,c\bar{c} V_{\mathrm{cl}}(i) |\Psi} = {\cal A}_\Psi^{\mathrm{disk}}(V_{\mathrm{cl}}) -  {\cal A}_0^{\mathrm{disk}}(V_{\mathrm{cl}})
\ee
is equal to the difference of the closed string one-point function between the vacuum described by the string field $\Psi$ and the reference perturbative vacuum $\psi=0$. He found that for the tachyon vacuum solution ${\cal A}_\Psi^{\mathrm{disk}}(V_{\mathrm{cl}})$ vanishes, in agreement with the fact that in the absence of D-branes there are no disk amplitudes to be computed. Since then it has become a standard practise to evaluate this invariant every time a new analytic solution appears, as in \cite{ErS}. One could also consider the full boundary state constructed in \cite{KOSZ} but for simplicity we restrict ourselves to the Ellwood invariant only.

Let us thus compute the invariant for a general solution of the form (\ref{Fsol}). By a simple modification of Ellwood's original computation or applying the scaling arguments of \cite{ErS} one easily derives
\bea
\aver{c\bar c V_{\mathrm{cl}}(i) c \tilde F B c F^2} &=& \int_0^\infty d\alpha \int_0^\infty d\beta \, \frac{2i}{\pi} \beta \tilde f(\alpha) f(\beta) \aver{V_{\mathrm{cl}}(i)}_{UHP}^\mathrm{matter}\nonumber\\
&=& \tilde F(0) \partial F^2(0) \, {\cal A}_0^{\mathrm{disk}}(V_{\mathrm{cl}}),
\eea
where $f$ and $\tilde f$ are inverse Laplace transforms of $F^2$ and $\tilde F$ respectively. For a solution of the equations of motion $\tilde F = K/(1-F^2)$ and hence
\be
\aver{c\bar c V_{\mathrm{cl}}(i) \, F c \frac{K B}{1-F^2} c F} =  - \lim_{z \to 0}\, z \frac{G'(z)}{G(z)} \, {\cal A}_0^{\mathrm{disk}}(V_{\mathrm{cl}}),
\ee
where $G=1-F^2$ as in the previous section. We have used cyclicity of the bracket and the fact that $F$ commutes with midpoint insertions. Interestingly, we see that for $G$ holomorphic or meromorphic at zero, the Ellwood invariant and hence also the new background one-point function is given by an integer multiple of the one-point function in the perturbative vacuum.

The closed string one-point function for an appropriate choice of the graviton vertex operator can measure the energy of the configuration, and therefore the Ellwood invariant would measure the difference in energy between the new configuration and that of the perturbative vacuum. We can thus write an alternative formula for the energy of the class of solutions (\ref{Fsol}) as
\be\label{resB}
\frac{E}{\rm Vol}  = -\frac{1}{2\pi^2} \lim_{z \to 0}\, z \frac{G'(z)}{G(z)},
\ee
where we fixed the overall coefficient by matching to the correct answer for the tachyon vacuum.

At first sight the results (\ref{resA}) and (\ref{resB}) look rather different. Note however that the function $G$ is required to be holomorphic in the right half plane. If we assume it is meromorphic at zero, then $G'/G$ can have at most first order pole and hence (\ref{resB}) can be rewritten as
\be\label{resB2}
\frac{E}{\rm Vol}  = -\frac{1}{2\pi^2} \oint_{C_0}\!\! \frac{dz}{2\pi i}\, \frac{G'(z)}{G(z)},
\ee
where $C_0$ is now a small contour around the origin. The difference between (\ref{resA}) and (\ref{resB2}) is given by a contour integral along the imaginary axis, bypassing possible singularity at zero on the right, and closing at infinity in the $\re z<0$ plane. Assuming that the function $G'/G$ happens to be holomorphic at infinity we can then shrink the new contour $C-C_0$ around the point at infinity and we get zero. This shows that under the above mentioned assumptions both  energy computations (\ref{resA}) and (\ref{resB}) give identical answer.

\section{Discussion}

We have shown that the class of string field solutions of the form
\be
\Psi = F c \frac{K B}{1-F^2} c F
\ee
have their energy density equal to
\be
\frac{E}{\rm Vol} = \frac{1}{2\pi^2} \oint_{C}\!\! \frac{dz}{2\pi i}\, \frac{G'(z)}{G(z)} = -\frac{1}{2\pi^2} \lim_{z \to 0}\, z \frac{G'(z)}{G(z)}
\ee
under certain holomorphicity assumptions. For holomorphic functions both expressions give integer multiples of single D-brane tension and hence a very natural guess would be that some of these solutions describe configurations of multiple D-branes. For $G(z) \sim z^n$ for small $z$ the energy comes out negative. Of course $n=1$ is the tachyon vacuum, but what about $n \ge 2$? This looks like a negative number of branes. We thus expect the solution to be sick and one of the signs of this sickness is a term of negative scaling dimension in (\ref{Fsol}) with dimension less than $-1$. This implies that at least some coefficients in the Fock basis will be divergent.

The really interesting question is whether solutions with positive number of branes can make some sense and it is a subject of an ongoing investigation \cite{progress}. The function $G$ has to behave as $G \sim z^{-n}$ and such a solution would describe a configuration of $n+1$ D-branes. The main problem is that such a $G$ is not a well defined $L_0$-safe string field and it is not yet clear whether this presents a substantial problem also for $\Psi$. Happily, the negative scaling dimension terms, which ruled out the negative tension configurations, are not present in this case. One possible example of multiple D-brane solutions would be given by
\be
G(z) = \left(\frac{z+1}{z}\right)^n,
\ee
but we have not yet verified whether the corresponding solution $\Psi$ has a well behaved level expansion. Additionally, we would like to see enhancement of cohomology around such solutions as well.

\section*{Acknowledgments}

\noindent
 We would like to thank the Yukawa Institute for Theoretical Physics at Kyoto University for hospitality during final stages of this work and for organizing the conference String Field Theory and Related Aspects where our work was presented. We would like to thank Ian Ellwood, Ted Erler and Ashoke Sen for useful discussion. M.S. would also like to thank the participants of the Aspen Center for Physics summer workshop for emphasizing the need of computing the energy of the solutions considered in this paper. The work of M.M. (No. 21-173) was supported by Grants-in-Aid for Japan Society for the Promotion of Science (JSPS) Fellows. The research of M.S. was supported by the EURYI grant GACR EYI/07/E010 from EUROHORC and ESF.

\end{document}